\documentclass[twocolumn,aps,prd]{revtex4-1}
\usepackage{graphicx}
\usepackage{bm}

\newcommand{\Y}{\Y_{lm}}

\newcommand{\be}{\begin{equation}}
\newcommand{\ee}{\end{equation}}
\newcommand{\Be}{\begin{eqnarray}}
\newcommand{\Ee}{\end{eqnarray}}

\newcommand{\f}{\frac}
\begin{document}
\pagestyle{plain}

\title{Flare-out condition of Morris-Thorne wormhole and finiteness of pressure}

\author{Sung-Won Kim}
\email[email:]{sungwon@ewha.ac.kr}
\affiliation{Department of Science Education, Ewha Womans University, Seoul 120-750, Korea}
\date{today}

\begin{abstract}
Wormhole is defined as the topological structure with the throat connecting two asymptotically flat spaces. In order to have and maintain the structure of the wormhole, there needs the geometrical flare-out condition, i.e., the minimal size at throat.
In the case of Morris-Thorne type wormhole, the condition is given by the huge surface tension compared to the energy density times the square of the light speed.
In this paper, we re-considered the flare-out condition for the wormhole with the Einstein equation, checked the finiteness of the pressure, and investigated its physical meaning.
\end{abstract}
\maketitle

The Lorentzian wormhole consists of two asymptotically flat spacetimes and the bridge connecting two spacetimes \cite{mtw}. In general, the flatness of two spacetimes is not necessarily required for constructing wormholes. There are several examples of the wormhole with non-flat spacetimes. The examples are the wormholes in de Sitter model \cite{roman} or FLRW (Friedmann-Lema\^{\i}tre-Robertson-Walker) cosmological model \cite{kim96}.
Among the models of wormhole spacetime, the Morris-Thorne type wormhole is widely used for treating the various physical issues of static wormhole  \cite{exam1,exam2,exam3,exam4,exam5,HKL,Lens}.
The spacetime is given as \cite{mt}
\be
ds^2 = e^{2\Phi(r)}dt^2 - \left( 1 - \f{b(r)}{r} \right)^{-1}dr^2 - r^2 (d\theta^2 + \sin ^2 \theta d \phi^2 ),
\ee
where $\Phi(r)$ is the redshift function and $b(r)$ is the wormhole shape function.
The flare-out condition is more understandable through the embedding geometry, as shown in Fig.~1. The embedded spacetime at $t=\mbox{const}$ and $\theta=\pi/2$ is
\Be
dS_2^2 &=& \left( 1 - \f{b(r)}{r} \right)^{-1} dr^2 + r^2 d\phi^2
= dz^2 + dr^2 + r^2 d\phi^2 \nonumber \\
&=& \left[ 1 + \left( \f{dz}{dr} \right)^2 \right]dr^2 + r^2 d\phi^2.
\Ee
\begin{figure}[b]
\includegraphics[height=4cm]{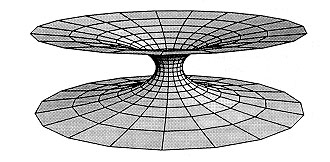}
\caption{The embedded shapes of the wormhole}
\end{figure}
Here
\be
\f{dz}{dr} = \pm \f{1}{\sqrt{r/b-1}} \quad \mbox{or}
\quad
\f{dr}{dz} = \pm \sqrt{\f{r}{b}-1}.
\ee
The flare-out condition in geometry is given by the minimality of the wormhole throat as
\be
\f{d}{dz} \left( \f{dr}{dz} \right) = \f{b-b'r}{2b^2} > 0.\label{geo-flare}
\ee
It is determined by $b$, independently of $\Phi$.
The most flare-out conditions have been presented as the minimal size of throat for clarifying the wormhole structure \cite{flare1,flare2,flare3,flare4,flare5,flare6,HMV,HV}.
 Hochberg and Visser \cite{HV} suggested the `simple flare-out condition' by adding the equality to the inequality in Eq.~(\ref{geo-flare}).
 Here, we discuss only the `strong flare-out condition' with the inequality like Eq.~(\ref{geo-flare}).
The minimality is reinterpreted as the divergent null rays through throat, which means the null energy condition violation from the Raychaudhuri equation \cite{mt}.
This flare-out condition can be rewritten in terms of wormhole matter.
The Einstein equation is
\Be
\rho &= & \f{b'}{8\pi r^2}, \\
\tau &=& \f{b/r - 2(r-b)\Phi'}{8\pi r^2}, \label{tau}\\
P &=& \f{r}{2}[(\rho-\tau)\Phi' - \tau' ] - \tau, \label{pressure}
\Ee
where $\rho$ is the energy density, $\tau$ is the surface tension, and $P$ is the pressure.
Here, the function $\rho$ is determined by $b(r)$, while $\tau$ is determined by $b(r)$ and $\Phi'(r)$.
We can rewrite $P$ in terms of $b$ and $\Phi$ as
\Be
P &=& \f{1}{8\pi r}\left[ (r-b) \left( \Phi''+(\Phi')^2 + \f{\Phi'}{r} \right)\right. \nonumber\\
&& \left.- \left( \f{b'r-b}{2r} \right) \left( \Phi' + \f{1}{r} \right) \right].
\Ee
 If we write $\Phi'$ in terms of $\tau$ and $b$, it becomes as
\be
\Phi' = \f{b-8\pi r^3 \tau}{2r(r-b)} \label{phiprime}
\ee
from Eq.~(\ref{tau}).
Usually the exoticity function $\zeta$ is used for the flare-out condition as
\Be
\zeta \equiv \f{\tau-\rho}{|\rho|} &=& \f{b/r -b'- 2(r-b)\Phi'}{|b'|} \nonumber \\
&=& \f{2b^2}{r|b'|}\left( \f{d^2r}{dz^2} \right) - \f{2(r-b)\Phi'}{|b'|}. \label{exoticity}
\Ee
In the Morris-Thorne paper \cite{mt}, the function $\zeta$ has the same sign as the first term of Eq.~(\ref{exoticity}), since the second term vanishes at or near throat $r=b$, so that the flare-out condition is replaced by \cite{mt,mty}
\be
\zeta = \f{\tau-\rho}{|\rho|} > 0 \label{flare-out}
\ee
at or near the throat as the flare-out condition in matter. Generally, the term $(\tau-\rho)$ near or at throat is determined by $b$ only, as we see in Eqs.~(\ref{geo-flare}) and (\ref{exoticity}). Here $\Phi'$ does not affect the value of $(\tau - \rho)$ near or at throat.
Thus it was claimed that the wormhole should have the large surface tension compared to the energy density to maintain the structure.
This condition seems be plausible physically. This violates the weak energy condition and sometimes violates the `averaged' weak energy condition to minimize the use of exotic matter \cite{mt}.

However, it seems that the flare-out condition Eq.~(\ref{flare-out}) cannot be accepted generally, because the prerequisite for it is 
\be
\lim_{r\rightarrow b} (r-b)  \Phi'=0 \label{zero}
\ee
which is guaranteed only for special value of $\tau$. As we see in Eq.~(\ref{phiprime}), $\Phi'$ has the $1/(r-b)$ term which might cancel the $(r-b)$ factor of second term in Eq.~(\ref{exoticity}) so that the limiting value of the term becomes non-zero.
If it is not zero, $\Phi'$ diverges in the limit of $r\rightarrow b$, still maintaining the continuity of $\Phi$. Apparently, it cannot be certain that the flare-out condition Eq.~(\ref{flare-out}) is the same as Eq.~(\ref{geo-flare}).

Now we should check the unused Einstein equation for $P$, Eq.~(\ref{pressure}),  whether the condition for $\Phi'$ Eq.~(\ref{zero}) is acceptable or not. The equation is useful for checking the finiteness of $P$ in the limit of   $r\rightarrow b$.
We will consider this problem by analyzing the term in the following two cases: $\lim_{r\rightarrow b} (r-b)\Phi' = 0$ and $\lim_{r\rightarrow b} (r-b)\Phi' \neq 0$.

In the first case, $\lim_{r\rightarrow b} (r-b)\Phi' = 0$, we can set $\Phi'$ as
\be
(r-b)\Phi' =  (r-b)^k A(r), \qquad k \geq 1
\ee
with finite (including zero) $A(r)$ in this limit. The tension $\tau$ is given as
\be
\tau = \f{b}{8\pi r^3} - \f{(r-b)^k}{4\pi r^2}A(r). \label{tau-zero}
\ee
Then $\Phi'$ and $\Phi''$ are
\Be
\Phi' &=& (r-b)^{k-1}A(r), \\
\Phi'' &=& (k-1)(r-b)^{k-2}A(r)+(r-b)^{k-1}A'(r) \\
&=& (r-b)^{k-1}\left[ \f{k-1}{r-b}A(r)+A'(r)\right].
\Ee
In the limit of $r\rightarrow b$, $\Phi'$ and $\Phi''$ are finite for any $k \ge 1$. As we see Eq.~(\ref{tau-zero}), $\tau$ has the finite value $b/8\pi r^3$ in the limit. We should check for the finiteness of $P$ at throat through the Einstein equation Eq.~(\ref{pressure}). By using the values of $\Phi'$ and $\Phi''$, we can show that $P$ will be finite for all values of $k \ge 1$. When $k=1$,
\be
\Phi'=A(r), \quad \Phi''=A'(r), \quad \tau=\f{b}{8\pi r^3} - \f{r-b}{4\pi r^2}A(r).
\ee

In the second case, $\lim_{r\rightarrow b} (r-b)\Phi' \neq 0$, $\Phi'$ can be rewritten as
\be
\Phi' = \f{B(r)}{r-b},
\ee
where $B(r) = (b-8\pi r^3\tau)/2r$ and has non-zero finite value in the limit.
Then
\be
\tau = \f{b}{8\pi r^3} - \f{B(r)}{4\pi r^2} \label{tau-nonzero}
\ee
and
\Be
\Phi' &=& \f{B}{r-b}, \label{phiprime_B}\\
\Phi'' &=& \f{B'}{r-b} - \f{(1-b')B}{(r-b)^2}. \label{phidoubleprime}
\Ee
In the limit of $r\rightarrow b$, $\Phi'$ and $\Phi''$ are not finite, while $\tau$ is finite in Eq.~(\ref{tau-nonzero}). To check the finiteness of $P$, we should examine the Einstein equation Eq.~(\ref{pressure}) with Eq.~(\ref{phiprime_B}) and Eq.~(\ref{phidoubleprime}) as
\Be
P &=& \f{1}{8\pi r}\left[ B'+\f{B}{r} - \f{b'r-b}{2r^2} \right. \nonumber \\
&& \left. + \f{B}{(r-b)} \left(
B - (1-b') - \f{b'r-b}{2r} \right) \right]. \label{P_B}
\Ee
There are two possible conditions for finite $P$, which can be obtained by vanishing the term including $(r-b)^{-1}$ in Eq.~(\ref{P_B}). One condition for finite $P$ in the limit is
\be
B - \left(1-b'+\f{b'r-b}{2r}\right) = 0
\label{condition1}
\ee
and another one is
\be
B - \left(1-b'+\f{b'r-b}{2r}\right) = (r-b)^\ell C(r), \qquad \ell \ge 1 \label{condition2}
\ee
where $C(r)$ is finite function of $r$ in the limit.  However, the first condition Eq.~(\ref{condition1}) is  unacceptable since it gives rise to the black-hole--like solution which should be excluded in treating the wormhole issues. In this condition Eq.~(\ref{condition1}), $\Phi$ and $g_{tt}$ become
\be
\Phi = \ln\left(r\sqrt{1-\f{b}{r}}\right) \quad \mbox{or}
\quad
e^{2\Phi} = r^2 \left(1-\f{b}{r}\right),
\ee
which shows the event horizon.
The second condition Eq.~(\ref{condition2}) make $g_{tt}$ as
\be
e^{2\Phi}=r^2\left(1-\f{b}{r}\right)e^{2\int(r-b)^{\ell-1}C(r)dr} \label{sol:26}
\ee
with given $\ell$ and $C(r)$. The special example as $\ell=1$ and $C(r)=1$ shows
\be
e^{2\Phi}=r^2\left(1-\f{b}{r}\right)e^{2r}
\ee
which does not change the nature of event horizon.
Any $\ell\ge 1$ and non-zero finite $C(r)$ in Eq.~(\ref{sol:26}) cannot cancel the term $(1-b/r)$.
Thus $\Phi$ given as Eq.~(\ref{condition2}) cannot be acceptable as the regular redshift function.
When the nonzero value of $\lim(r-b)\Phi'$ diverges, that is, $\lim B(r)$ diverges, then $B(r)$ has $\ln(r-b)$ or $(r-b)^{-m}, m \ge 1$ containing term. With these values of $B$, $P$ also diverges in the limit. The results show that
we cannot find finite $P$ in the second case,  $\lim (r-b)\Phi' \neq 0$. Thus this second case is not allowed for defining the flare-out condition of the Morris-Thorne wormhole.

Therefore, we conclude that the flare-out condition which guarantees the finiteness
$\Phi'$ and other physical quantities at throat is come from the condition of $\lim (r-b)\Phi' = 0$ only. The condition Eq.~(\ref{flare-out}) is equal to the condition Eq.~(\ref{geo-flare}).

To examine the problem in more practical way,
we will consider the properties of $\Phi(r)$ and $\Phi'(r)$,
and re-derive the flare-out condition in case of the power-law distributions of $\rho$ and $b$.

Now we check the properties of $\Phi$ and $\Phi'$. There are some restrictions on the redshift function $\Phi$ to prevent any event horizon. Since there are no event horizon, $\Phi$  must be everywhere finite \cite{mt}. On account of the asymptotically flatness of the wormhole, $\lim_{\ell\rightarrow\pm\infty}\Phi(\ell)=\Phi_\pm$. And $\Phi$ is continuous across the throat, i.e., $\Phi_+(r_0) = \Phi_-(r_0)$.
The condition of $\Phi'$ is \cite{visser}
\be\Phi'_+(r_0) = \Phi'_-(r_0), \label{cond_phiprime}
\ee
which is derived from the condition Eq.~(\ref{zero}). In this condition, $\Phi$ do not allow any divergencies in the limit.

As an alternative way, we treat the flare-out condition of the Morris-Thorne wormhole in the case of the power-law distribution of matter.
If the energy density $\rho$ is given by
\be
\rho = \rho_0 \left( \f{r}{r_0} \right)^n,
\ee
the condition Eq.~(\ref{geo-flare}) becomes $n < -2$. Meanwhile, if the shape function is given as $b=b_0(r/r_0 )^q$ instead of $\rho$, the condition is
\be
\f{d}{dr}\left( \f{b}{r} \right) < 0,
\ee
or $q<1$. This two power indices $n$ and $q$ have the relationship of $q=3+n$.
Therefore, we can say that the flare-out condition is $n<-2$ for $\rho = \rho_0(r/r_0 )^n$, necessarily.
If we do not know the detailed form of $\Phi'$, we can't decide $\tau(r)$, since
 the tension $\tau$ is determined by $\Phi'$ and $b$.  The flare-out condition is determined by $b$ and $b'$, that is, $\rho(r)$ can decide the condition, such as $n<-2$. The tension $\tau$ does not play direct role in presenting the flare-out condition.
If $\Phi'$ is restricted in some regions, $\tau$ can be restricted.

In the special simplest case, $\Phi=0, b=b_0^2/r$, Einstein equation is
\be
\rho = \f{b'}{8\pi r^2} = - \f{b_0^2}{8\pi r^4} < 0.
\ee
Since $n=-4 < -2$ here, this distribution of matter satisfies the flare-out condition.
The geometric flare-out condition Eq.~(\ref{geo-flare}) is
\be
\f{b-b'r}{2b^2} = \f{r}{b_0^2} = \f{1}{b} >0.
\ee
It automatically satisfies the flare-out condition except the requirement of negative energy density. The other matter terms are
\Be
\tau &=& \f{b}{8\pi r^3} = \f{b_0^2}{8\pi r^4} > 0, \\
P &=& \f{r}{2}(-\tau')-\tau = \f{b_0^2}{8\pi r^4} = \tau.
\Ee

There are other models for wormhole using exotic matter. We will check what the flare-out condition is for each cases.

For the delta function type wormhole, the minimality is trivial. Since the matters are given as the negative value
\be
\sigma=-\f{1}{2\pi b}, \quad \vartheta = - \f{1}{4\pi b},
\ee
the flare-out condition is
\[
\f{d^2r}{dz^2} = \f{b-b'r}{2b^2} = \f{1}{2b} > 0.
\]
The delta function-type wormhole also satisfies the flare-out condition.
Similarly, the thin-shell models also satisfy the condition.

To summarize, we reconsidered the flare-out condition of the wormhole.
The flare-out condition is given by the energy density only, since the shape function is determined by the energy density. The shape function has the relationship with tension and $\Phi'$.
However, the finiteness of $P$ restrict the value of $\Phi'$ in the limit of  $r\rightarrow b$ for the case of Morris-Thorne wormhole. This shows that the high tension flare-out condition is reasonable and acceptable without any flaw.
Also this condition can be replaced by the condition of the power indices for power-law model.

We deeply feel thanks to Prof. F. Lobo for helpful discussions.
This research was supported by Basic Science
Research Program through the National Research Foundation of Korea
(NRF) funded by the Ministry of Education, Science and Technology
(2010-0013054).


\end{document}